\documentstyle[12pt]{article}

\begin{document}

\title{
Resonance Production on Nuclei at High Energies:\\
Nuclear-Medium Effects and Space-Time Picture}
\author{K.G. Boreskov and L.A. Kondratyuk \\[2mm]
{\small {\it Institute of Theoretical and Experimental Physics,}}
\\[1mm]{\small\it Moscow, Russia} \\[2mm]
M.I. Krivoruchenko\thanks{
Permanent address: Institute for Theoretical and Experimental Physics,
B. Cheremushinskaya 25, 117259 Moscow, Russia} \\
{\small {\it Institut f\"ur Theoretische Physik, Universit\"at
T\"ubingen,}}\\[1mm]
{\small {\it T\"ubingen, Germany}} \\[2mm]
{ and}\\[2mm]
J.H. Koch \\[2mm]
{\small {\it National Institute for Nuclear Physics and High Energy
Physics (NIKHEF)}}\\[1mm]
{\small {\it and Institute for Theoretical Physics, University of 
Amsterdam,}}\\[1mm]
{\small {\it Amsterdam, The Netherlands}}}
\date{}

\maketitle

\begin{abstract}
The influence of nuclear matter on the properties of coherently
produced resonances is discussed. It is shown that, in general,
the mass distribution of resonance decay products has a
two--component structure corresponding to decay outside and
inside the nucleus. The first (narrow) component of the amplitude 
has a Breit--Wigner form determined by the vacuum values of mass
and width of the resonance. The second (broad) component
corresponds to interactions of the resonance with the nuclear medium. 
It can be also described by a Breit--Wigner shape with parameters 
depending {\it e.g.} on the nuclear density and on the cross section of 
the resonance--nucleon interaction.

The resonance production is examined both at intermediate energies,
where interactions with the nucleus can be considered as a series of
successive local rescatterings, and at high energies,
$E>E_{crit}$, where a change of interaction picture occurs.
This change of mechanisms of the interactions with the nucleus is
typical for the description within the Regge theory approach and is 
connected with the nonlocal nature of the reggeon interaction.
\end{abstract}

\newpage


\section*{\bf\large 1.INTRODUCTION}
\indent

The study of the propagation of hadrons in nuclear matter is a
much studied subject. It is a tool to explore the structure of
the nucleus and to probe nuclear matter at large densities.
Another perspective, which we adopt here, is to consider the
nucleus as a known microscopic laboratory for studying elementary
hadronic properties and interactions, which can then be compared
to the predictions of models.  This is particularly important in
the case of unstable hadrons (resonances), where nuclear
measurements are the only means to study their interaction.  Even
when the free elementary process is well known, experiments on a
nucleus are necessary to tell us something about the space--time
picture of the elementary interaction by providing a measure --
typical nuclear distances and time scales.  The bulk effect of
the target nucleons on the propagation of the hadron is often
referred to as 'medium modification' of the particle properties,
expressed through an 'in-medium' mass and, for a resonance, also
width of the hadron. 

The creation and propagation of resonances in nuclei has been widely
studied over the last decades, both experimentally and theoretically.
There are two distinct ways to create a resonance in the
collision of a projectile with a nuclear target. In the first,
the projectile and a target nucleon form a resonance. This
$s$-channel resonance occurs at a definite projectile energy
and shows up as a peak in the energy dependence of the total
cross section. An example is the excitation of nucleon resonances,
in nuclear photoabsorption, where {\it e.g.} for the $\Delta$ resonance
the elementary process is $\gamma+N\rightarrow\Delta$.
It is common to refer to this type of resonance creation as
{\it formation}. Recent examples are the nuclear photoabsorption measurements 
at Frascati \cite{Fras} and Mainz \cite{MAMI}. They showed that some nucleon resonances that were 
seen in photoabsorption on hydrogen and deuterium dissapear in nuclei, which
to a large extent may be due to collision broadening of the resonance.
In the second type, called resonance {\it production}, other particles besides the resonance emerge after
the initiating projectile--nucleon interaction. For example in 
the nuclear photoproduction of the $\rho$ resonance, the underlying mechanism  is $\gamma+N\rightarrow\rho+N$.
There is no characteristic energy dependence in the cross section in this case and the resonance
has to be identified through the invariant mass of its decay
products. The production of a particular resonance in a given
process continues to play a role at very high projectile energy,
while {\em formation} of that resonance is suppressed as soon as
one is more than a half width removed from the resonance energy.
This is the reason that formation is essentially a phenomenon
relevant at low 'resonant' energies, while production is important
at high energies as well. 
A recent experiment studying the production of the $\rho$ meson is the 
measurement of di-lepton spectra at the SPS \cite{SPS1} -  \cite{SPS3}. The
results were interpreted through a shift of the $\rho$ mass in nuclear
matter, as predicted in effective Lagrangian models and in approaches
based on QCD sum rules.

Many theoretical treatments of the
behavior of resonances in nuclei were done for nuclear matter and then
applied to finite nuclei by taking over the 'in medium'
propagator. A central topic of this paper is the appropriate Green
function for finite nuclei. The finiteness of nuclei was also a crucial
ingredient in the work of Ericson \cite{Eri} on the propagation of virtual pions in nuclei
and our discussion has some common points.

We discuss in this paper the {\it coherent} diffractive production of 
hadronic
 resonances on nuclei such as the photoproduction of vector 
mesons.
In this case a resonance $R$ is produced by the initial particle $a$
on one of target nucleons, $a N\rightarrow RN$, propagates in
nuclear matter interacting with other nucleons, $RN\rightarrow RN$,
and then decays into a system of particles, $R\rightarrow
x_1,x_2,\ldots$. The decay products are registered {\em outside}
the nucleus, but the decay itself can occur both inside or
outside the nucleus. For simplicity, we will leave out final
state interactions of the decay particles, which would be appropriate
for the decay of a $\rho$ resonance into an $e^+ e^-$ pair.
It will be shown that in the case of {\em
coherent} production, where the nucleus returns to its ground state
after the reaction, the mass distribution of decay products
contains information on both stages of resonance life,
in nuclear matter and in the vacuum. As a result, the mass
distribution has to be interpreted not in terms of a single 'in medium' 
resonance, but as an amplitude with
a two--component structure. These two contributions can separately be parametrized through Breit--Wigner shapes. The first (narrow) peak corresponds
to the decay  outside the nucleus and is determined by the vacuum 
values of mass and width of the resonance. The second -- typically much
broader -- peak is due to the interaction  of the resonance with the 
nucleus and can be characterized through changed values of resonance 
mass and width. Clearly, the relative contribution of the two components 
depends {\it e.g.} on the nuclear density, the atomic number $A$
of the target or the life time of the resonance.
The interference of these two contributions gives the mass distribution of
the decay products a more complicated form and makes it harder to
extract the medium modified resonance parameters from the obseved cross 
section.

The most common way to think about the propagation of a hadron
in a nucleus
originated from the intermediate energy hadron--nucleus scattering.
At energies where the projectile wavelength is small compared
to the characteristic sizes of the process, the range of the
potential and the nuclear dimensions,
the process is seen as a succession of individual elementary
projectile--nucleon interactions, described by the free amplitude.
In particular at higher energies the motion of the nucleons can be
neglected, leading to the frozen nucleus picture as embodied in {\it 
e.g.} the often used eikonal description.

One might expect that the inherent approximations become better
with increasing energy. As was pointed out by Gribov
\cite{Gribov},\cite{GribovA} and followed up in detail by Koplik and Mueller
\cite{Koplik}, this is not the case.
For a high energy projectile above a critical energy,
$E>E_{crit}$, the picture changes radically. The fast
projectile should be seen as developing into a superposition of 
multiparticle hadronic states before reaching the target and before the first 
interaction takes place. That interaction takes place only between the
lowest energy component of the projectile and the target. For an
elastic scattering process, this multiparticle intermediate
state eventually recombines into the original projectile--target
state. This corresponds to an interaction of a considerable
nonlocality, which makes the traditional approximations leading
to the eikonal description incorrect. However, it was
shown \cite{Gribov},\cite{GribovA} that, by consistently incorporating all intermediate multiparticle
states in the coherent scattering of 
a stable hadron from a nucleus, one surprisingly enough obtains an
expression that has the same structure as the result obtained from a
naive application of the eikonal approach.
Here we extend these ideas to the production and
propagation of a resonance in a coherent production process on a nucleus,
such as the electromagnetic production of vector mesons.  It is
clearly also relevant to the studies of dense nuclear matter
through resonance production in relativistic nuclear reactions. 

The paper is organized as follows. In Chapter 2 we discuss
the production of resonances at intermediate energies, less than the 
critical energy, but high enough to use the eikonal approximation.
We first give a simple, qualitative explanation why 
the production amplitude in a finite nucleus must have a two-component 
structure. We then derive the full expression in the eikonal approximation 
and show that it displays a two--component structure for the invariant 
mass distribution of the decay products.
In Ch. 3, the situation for energies higher than the critical 
value, $E>E_{crit}$, is discussed. Using Gribov's method we show that the
final formulas turn out to have the same structure as for the eikonal
approach in spite of the completely
different space--time picture of the interactions in the two energy
regions. In case that only one reaction channel is relevant, the propagating 
resonance, the formulas are identical. The reader mainly interested in the
two-component structure of resonance production in finite nuclei 
could therefore go
from Ch. 2 straight to Ch. 4, where we give some numerical examples for 
the photoproduction of $\rho$ mesons on nuclei to show the relative importance 
of the two resonance components. The dependence on the nuclear size is studied 
and the influence of nuclear correlations on medium modified resonance 
parameters are considered. A summary and some conclusions are given in 
Chapter 5. A preliminary version of this work has been
presented in Ref. \cite{NAN}.

\section*{\bf\large 2. INTERMEDIATE ENERGIES}
\indent

In this section we consider the coherent production of resonances on 
a nucleus at energies less than $E_{crit}$, where the interaction
of the resonance with the nucleus can be described through
the multiple scattering of the resonance from the target nucleons,
summarized in an optical potential. The propagation of the resonance 
in the nuclear medium is then given by the corresponding 'dressed' 
Green function.

Before discussing this in detail, we first derive the relevant 
features of the production amplitude in a qualitative fashion
for a one-dimensional example, similar to the arguments used in Ref. 
\cite{Eri}.
We consider a high energy photon incident along the positive z axis 
with momentum $p$ that strikes an infinite slab of nuclear matter 
extending from 
$ - z_A$ to $z_A$. Inside the nuclear matter at $z_i$ a $\rho$-resonance
is excited. 
By matching the logarithmic derivatives at $z_A$ and neglecting backward 
motion of the resonance, a good approximation
at high energies, it is simple to see that the wavefunction of 
the resonance has the $z$ - dependence
\begin{equation}
\Psi_{z_i} (z) = N(z_i) { exp [ i\; P^{*} (z_A - |z|) ] \theta ( z - z_A) 
+  exp [ i\;P (z - z_A) ] \theta ( z - z_A) }\;
\label{eq:wavef}
\end{equation}
where $ N(z_i)$ is a factor depending on the excitation strength at $z_i$.
We assume that the influence of the nucleus on the resonance can be 
represented through a complex optical potential, $V^*$. Then $P^*$, 
the momentum of the resonance inside the nuclear medium, is 
\begin{equation}
P^* = \sqrt{p^2 - M_R^2 + i M_R \Gamma + 2 M_R V^*}\;
\equiv p - Q^*_R\;.
\end{equation}
where $M_R$ and $\Gamma_R$ are the free resonance mass and width,
respectively. The free resonance momentum outside the nucleus is given by
\begin{equation}
P = \sqrt{\tilde{s} - M_R^2 + i M_R \Gamma} \equiv p - Q_R\;.
\end{equation}
The resonance decays at point $z_f$ into an electron-positron pair,
a plane wave state of momentum $p-q$. The production amplitude 
is then simply the integral over excitation and decay points,
\begin{eqnarray}
&&T(p,q) =
\int_{-z_A}^{z_A} dz_i \int_{z_i}^{\infty} dz_f 
e^{-i(p-q)z_f}\Psi(z_f) \nonumber \\
&&\propto \int_{-z_A}^{z_A} dz_i N(z_i) \left\{
{1-\exp[i2pD^{\ast}(z_A-z_i)]\over D^{\ast}}
+{\exp[i2pD(z_A-z_i)]\over D} \right\} ~,
\label{eq:approxt}
\end{eqnarray}

where
\begin{equation}
2\,p\,D^* = q - Q_R^*\;,\;\;2\,p\,D = q - Q_R\;.
\end{equation}
For large $p$ we have $q \simeq \tilde{s}/2p$, where $\tilde{s}$ is
the total energy of the resonance. We can then write the denominators 
in Eq. (\ref{eq:approxt}) as
\begin{eqnarray}
D &\simeq & \tilde{s} - M_R^2 + i M_R \Gamma_R \;,\\
D^* &\simeq & \tilde{s} - {M^*_R}^2 + i M^*_R \Gamma^*_R\;.
\end{eqnarray}
At this stage, we can already read off the relevant features. First,
the amplitude is the sum of {\it two} Breit-Wigner resonance contributions,
consisting of a term with the free and medium modified denominators, $D$ and
$D^*$, respectively. There is no true pole in the complex $\tilde{s}$ plane
due to the in medium denominator. Furthermore, it can be seen that only in the 
limit of infinite nuclear matter, $z_A \rightarrow \infty$, where the
exponential terms vanish, we are left with an amplitude that has a 
single, medium modified component and a pole in the complex $\tilde{s}$ plane.
These are the main results we derive in this Chapter.

We now turn to a more detailed and complete discussion. Rather than working 
directly within the well eikonal formalism, we briefly rederive it 
from a diagrammatic approach. This useful in order to better connect 
to the discussion in Chapter 3. We start with the amplitude for resonance 
production by
a projectile $a$ with momentum $p$ striking a nucleus of mass $A$,
neglecting its further interactions with the nucleus.  
It corresponds to the diagram of Fig.1 and the amplitude has the form
\begin{equation}
\label{eq:1}
T_0(p,q^2,{\tilde s})=
t (R\rightarrow X)\,
G_{0}({\tilde s})~t_{0}(s,q^2)~F_A(q)
 ~,
\end{equation}
Here $t_{0}(s,q^2)$ is the amplitude for resonance production on a
free nucleon, $G_0(\tilde {s})$ is the propagator of the resonance
and $t (R\rightarrow X)$ is the amplitude of its decay
into the final state $X=\{x_1,x_2,\cdots\}~$.  The momentum transfer to
the nucleus equals $q$ and the invariant mass of the resonance is
denoted by $\tilde {s}=(p-q)^2$. We assume that the energy is high enough that 
Fermi motion can be neglected. Therefore, we use in the amplitude $t_0$ the 
average total energy of the projectile and 
struck target nucleon, $s=(p+P_A/A)^2$, where $p$ and $P_A$ are four-momenta of
the projectile and the nucleus, respectively. In principle $t_{0}$ also 
depends on the invariant mass of the resonance, $\tilde {s}$.
The main dependence of $T_0$, Eq. (\ref{eq:1}), on $\tilde {s}$ comes from 
the Green function $G_{0}(\tilde {s})$
and we therefore ignore the weaker dependence in $t_0$; similar considerations
are also applied to $t (R\rightarrow X)$. We will only
explicitly indicate the $s$ dependence when it is necessary to
avoid confusion.

We choose the $z$-axis along the beam direction. For coherent production
we obtain the following kinematical conditions, typical for the eikonal 
approximation at high energies:
\begin{eqnarray}
\label{eq:kine}
|q^2|\approx {\bf q}^2\sim R_A^{-2},
\nonumber \\
q_0\sim{\bf q}^2/2M_A\ll|{\bf q}|,
\nonumber \\
q_z \approx (\tilde {s}-m_a^2)/2|{\bf p}|,
\nonumber \\
|{\bf q}_{\perp}|\leq R_A^{-1}.
\end{eqnarray}

The nuclear form factor $F_{A}(q)$ is the Fourier transform
of the nuclear density,
\begin{equation}
\label{eq:2}
F_A (q)=\int d^3r\,e^{i{\bf q\cdot r}}\, \rho_A (r) ~.
\end{equation}

The propagator of the resonance,
\begin{equation}
\label{eq:3}
G_{0}(\tilde {s})=i\,[(p-q)^2 -M_R^2 +iM_R \Gamma_R ]^{-1} ~,
\end{equation}
with the above kinematical approximations and neglecting the $q^{2}$ term,
becomes
\begin{equation}
\label{eq:4}
G_{0}(\tilde {s})
=i\,\left[ 2|{\bf p}|\left( q_z -q_R +i\gamma_R/2\right) \right] ^{-1}
={i\over 2|{\bf p}|(q_z -Q_R)}~.
\end{equation}
Here 
\begin{equation}
\label{eq:4a}
q_R={M_R^2-m_a^2\over 2|{\bf p}|} ~,
~~~\gamma_R=\Gamma_R {M_R\over |{\bf p}|}
\end{equation}
are the minimal momentum transfer to produce the resonance $R$ and the Lorentz-factor reduced width of the resonance, respectively, 
which we combine into the complex quantity
\begin{equation}
\label{eq:4c}
Q_R=q_R-i\gamma_R/2~.
\end{equation}

We will in the following take the point of view of time ordered
perturbation theory, where the intermediate propagation of the
resonance occurs at fixed energy, 
\begin{equation}
\label{eq:ez}
E_0\approx |{\bf p}|+m_a^2/2|{\bf p}|,
\end{equation}
determined by the initial projectile momentum. 
It will be most convenient to work in the coordinate representation.
We therefore Fourier 
transform with respect to the momentum variables and obtain for propagation by
the distance ${\bf r} =(z, {\bf b})$ with ${\bf b \cdot p} = 0$:
\begin{eqnarray}
\label{eq:5}
G_0(z,{\bf b};E_0) &=& (2\pi)^{-3}\int dq_z\, d^2 q_{\perp}
\,e^{ipz-iq_z z-i{\bf q_{\perp} \cdot b}}\,G_{0}({\bf p}-{\bf q};E_0) 
\nonumber \\&=& 
e^{ipz}g_0 (z)\,\theta (z)\,\delta^2 ({\bf b}) ~,
\end{eqnarray}
where
\begin{equation}
\label{eq:6}
g_0 (z)=
{1\over 2|{\bf p}|}e^{-iq_R z-\gamma_R z/2}
=(2|{\bf p}|)^{-1} e^{-iQ_R z} ~.
\end{equation}
This Green function
has the natural form for a high energy process: forward propagation of the
resonance at fixed impact parameter ${\bf b}$ with an attenuation
proportional to $\gamma_R$.
The production amplitude can be written in a way reflecting the
space--time sequence of the process (with the initial state on the right):
\begin{equation}
\label{eq:7}
T_0(p,q^2,{\tilde s})=t (R\rightarrow X)\!
\int\! d^{3}r_f\!\int\! d^3 r_i\,
e^{-i({\bf p - q) \cdot r}_f}\,\,
G_0({\bf r}_f-{\bf r}_i;E_0)U_0({\bf r}_i)~e^{i{\bf p \cdot r_{i}}} 
 ~,
\end{equation}
where
\begin{equation}
\label{eq:8}
U_0({\bf r})=(2\pi)^{-3}\int d^3q \,e^{i{\bf q \cdot r}}\, t_0(q^2,s)\,F_{A}(q) ~. \\
\end{equation}
Eq.(\ref{eq:7}) describes the resonance production on a nucleon at the
point ${\bf r}_i=(z_i,{\bf b})$, its free propagation 
at constant impact parameter ${\bf b}$
to the point ${\bf r}_f=({z}_f,{\bf b})$ and its decay at
this point.

The corresponding cross section has the form
\begin{equation}
\label{eq:9}
d\sigma =\left| t (R\rightarrow X)\right|^2\,
{\left| t_{0}(s,q^2)\right|^2\over (\tilde {s} -M_R^2)^2 +M_R^2\Gamma_R^2}\,\,
{dq^2\,d\tilde{s}\over 32\pi^2 p\sqrt{s}}\,d\tau_X (\tilde {s})~.
\end{equation}
where $d\tau_X$ is the phase volume for the decay
$R\rightarrow X$.
One can see that the distribution $d\sigma /dq^2\,d\tilde {s}$ has the
usual Breit--Wigner form as a function of $\tilde {s}$.

What happens now when we allow for interactions
of the resonance with the nucleus?
One might simply expect that one again obtains a Breit--Wigner
distribution, but now for an 'in medium' resonance with modified
parameters $M_{R}$ and $\Gamma _{R}$. Below, we will show that this
is not the case for production of a resonance on a finite nucleus. 

To take into account the final state interactions of the resonance $R$ with
the nucleus (see Fig.2), we replace the free Green function, $G_0$ in Eq.(\ref{eq:7})
by the dressed Green function, $G$, which takes into account the interactions 
with the residual nucleus. 
The production amplitude obtained with the full Green function, $G$, becomes:
\begin{equation}
\label{eq:19}
T(p,q^2,\tilde{s})= t(R\rightarrow X) 
\int\! d^3 r_f \!\int\! d^3 r_i \,
e^{-i({\bf p- q) \cdot r}_f}\,G({\bf r}_f,{\bf r}_i;E_0)\,U_0({\bf r}_i)~
e^{i{\bf p \cdot r_{i}}} ~.
\end{equation}
For the construction of the dressed Green function 
in the usual multiple scattering picture, we use the elastic resonance--nucleon 
scattering amplitude, $t_R (s,q^2)$, normalized according to
\begin{equation}
\label{eq:13}
2\,{\rm Im}\,\ t_R (s,0) =\sigma_R^{tot}(s)~.
\end{equation}
Taking into account that the range of strong interactions 
is small compared to the nuclear size one can write approximately 
for the corresponding 'optical potential' that describes the interaction
with the nucleons in the nuclear ground state 
\begin{equation}
\label{eq:14}
U_R({\bf r})=(2\pi)^{-3}\!\int\! d^3q \,e^{i{\bf q \cdot r}}\,t_R(q^2, s)
F_A (q^2) \simeq
t_R (s,0)\,\rho_A({\bf r}) ~.
\end{equation}
For much of the discussion below the precise form of the interaction
does not matter. Other effects that vary approximately as the
nuclear density and can be included analogously will lead to similar 
conclusions.

In coordinate space the dressed Green function has the form
\begin{equation}
\label{eq:15}
G({\bf r}_f,{\bf r}_i;E_0)
=e^{ip(z_f-z_i)}g({z}_f,z_i)\,
\theta({z}_f-z_i)\,\delta^{(2)}({\bf b}_f-{\bf b}_i) ~,
\end{equation}
with
\begin{eqnarray}
g({z}_f,z_i) &=&
\sum_{n=0}^{\infty}i^n
\!\int\!\cdots\!\int_{{\cal Z}} dz_n\,\ldots \,dz_1\,
g_0({z}_f-z_n)\,U_R(z_n,{\bf b})
\nonumber \\
&&\times g_0(z_n-z_{n-1})\cdots
U_R(z_1,{\bf b})\,g_0(z_1-z_i)
\nonumber \\
&=& (2|{\bf p}|)^{-1}\exp [-iQ_R(z_{f}-z_{i})]
\sum_{n=0}^{\infty}
{1\over n!}\left[i\!\int_{z_i}^{z_f} dz~\! 
U_R(z,{\bf b})\right]^n
\nonumber \\ 
\label{eq:15a}
&=& (2|{\bf p}|)^{-1}\exp \left[ -iQ_R(z_f-z_i) 
+i\!\int_{z_i}^{z_f} dz\, U_R(z,{\bf b})\right] ~, 
\end{eqnarray}
and where the integrations over $z_i$'s are longitudinally ordered, {\it i.e.} 
\begin{equation}
\label{eq:17}
{z_i<z_1<z_2\cdots <z_n<z_f}~.
\end{equation}

It can easily be seen from Eqs.(\ref{eq:15}) and 
(\ref{eq:15a}) that for homogeneous
infinite nuclear matter with density $\rho_{0}$, the full Green function
$G({\bf r}_f,{\bf r}_i;E_0)$ depends only on the difference
${\bf r}_f-{\bf r}_i$.
This is due to translational invariance and yields momentum
conservation in the multiple scattering process. 
The full Green function can in this case be obtained from the free one,
Eqs.(\ref{eq:5}) -- (\ref{eq:6}), by the replacement
\begin{equation}
\label{eq:21}
Q_R\rightarrow Q_R^{\ast}
=Q_R-t_R(0)\rho_0 ~,
\end{equation}
or
\begin{equation}
\label{eq:replace}
M_R\rightarrow M_R^{\ast} ~,~
\Gamma_R\rightarrow\Gamma_R^{\ast}~,
\end{equation}
where
\begin{eqnarray}
\label{eq:22} 
M_R^{\ast 2} &=& M_R^2-2|{\bf p}|\,{\rm Re}\,t_R(0)\rho_0 ~, \\
\label{eq:23}
M_R^{\ast}\Gamma_R^{\ast} &=& M_R\Gamma_R+2|{\bf p}|\,{\rm Im}\, t_R(0)\rho_0
\equiv\gamma_R^{\ast}\,|{\bf p}| ~,~~~
\gamma_R^{\ast}=\gamma_R+ \sigma_R\,\rho_0 ~. 
\end{eqnarray}
For {\it infinite} nuclear matter, we thus obtain 
a production amplitude, $T$, for which the ${\tilde s}$ dependence is 
again given by a single Breit-Wigner denominator as in
Eq.(\ref{eq:9}), but now with the medium modified parameters $M_R^{\ast}$ 
and $\Gamma_{R}^{\ast}$. From Eq.(\ref{eq:23}) it is clear that the
nuclear matter width of the resonance is larger than the free one,
yielding a broader peak of the production cross section. The 
resonance peak also shifts to a different position, but whether this
shift is repulsive or attractive depends on the sign of ${\rm Re}\,t_{R}$.

The ${\tilde s}$ dependence of the cross section
for a {\it finite} nucleus is already revealed by carrying out the 
integration over the final coordinate, ${\bf r}_f$, in Eq.(\ref{eq:19}),
which amounts to taking the partial Fourier transform 
of the Green function $G({\bf r}_f,{\bf r}_i;E_0)$ 
\begin{equation}
\label{eq:20}
G({\bf r}_{i};{\bf p-q};E_0)=\!\int\! d^3r_f\,
e^{-i({\bf p-q}){\bf r}_f}\,
G({\bf r}_{f},{\bf r}_{i};E_{0}) ~.
\end{equation}
This function describes the amplitude to find the resonance $R$, after having
been produced at the point ${\bf r}_i$, in a plane wave
final state with momentum ${\bf p}-{\bf q}$ and with an invariant
mass 
\begin{equation}
{\tilde s}=
E_0^2-({\bf p - q})^2\approx 2|{\bf p}| q_z+m_a^2.
\end{equation}

For simplicity, we now assume that we are dealing with a nucleus of
radius $R_A$ and constant nuclear density, $\rho_0$ 
\begin{equation}
\label{eq:24}
\rho_A({\bf r})=\rho_0\,\theta(R_A-r)~.
\end{equation}
When Fourier transforming the Green function, Eq.(\ref {eq:20}), one receives 
two contributions corresponding to the decay inside 
and outside the nucleus, $r_f < R_A$ and $r_f > R_A$, respectively. 
As can be 
seen from Eq. (\ref{eq:15a}), these contributions can be written in the
simple form
\begin{equation}
\label{eq:25}
G( {\bf r}_i;{\bf p} - {\bf q};E_0)=C_{in}\cdot 
G_{in}({\bf r}_i;{\bf p} - {\bf q};E_{0})
+C_{out}\cdot G_{out}({\bf r}_i;{\bf p} - {\bf q};E_0) ~,
\end{equation}
where 
\begin{eqnarray}
\label{eq:26}
G_{in}({\bf r}_i;{\bf p} - {\bf q};E_0) &=&
i\,\left[2|{\bf p}|(q_z-Q_{R}^*)\right]^{-1} ~,
\nonumber \\
G_{out}({\bf r}_i;{\bf p} - {\bf q};E_0) &=&
i\,\left[2|{\bf p}|(q_z-Q_{R})\right]^{-1} ~.
\end{eqnarray}
The coefficients in Eq.(\ref{eq:25}) are given by
\begin{eqnarray}
\label{eq:27a}
C_{in}({\bf r}_i;{\bf p} - {\bf q},E_0) &=& 
\exp (-i(p-q_z)z_i)
\left\{1-\exp [i(q_z-Q_{R}^{*})(z_A-z_i)]\right\}
\nonumber \\
C_{out}({\bf r}_i;{\bf p} - {\bf q},E_0) &=& 
\exp (-i(p-q_z)z_i) \exp [i(q_z-Q_{R})(z_A-z_i)] ~,
\end{eqnarray}
where $z_A=\sqrt{R_A^2-{\bf b}^2}$ defines the point where the resonance
with impact parameter ${\bf b}$ leaves the nucleus.

To explicitly display the ${\tilde s}$ dependence and singularity structure,
we re-write this by using the kinematics of the eikonal 
approximations, Eqs.(\ref{eq:4} - \ref{eq:4a}) and the definitions 
in Eqs.(\ref{eq:21})--(\ref{eq:23}):
\begin{eqnarray}
\label{eq:28}
G_{in}(z_i,{\tilde s}, E_{0}) &=& 
\left [{\tilde s} -M_R^{*2} +iM_R^* \Gamma_R^* \right ]^{-1} ~, 
\nonumber\\
G_{out}(z_i,{\tilde s}, E_{0}) &=& 
\left [{\tilde s} -M_R^2 +iM_R \Gamma_R \right ]^{-1} ~,
\end{eqnarray}
and
\begin{eqnarray}
\label{eq:plus}
C_{in}({\bf r}_i;{\bf p}-{\bf q},E_0) &=&
\exp (-i(p-q_z)z_i) [1- \exp (\frac{i}{2|{\bf p}|}~({\tilde s} - 
M_R^{*2} +iM_R^* \Gamma_R^*)(z_A-z_i)] ~,
\nonumber\\
C_{out}({\bf r}_i;{\bf p}-{\bf q},E_0) &=&
\exp (-i(p-q_z)z_i) [\exp (\frac{i}{2|{\bf p}|}~({\tilde s} -
M_R^{2} +iM_R \Gamma_R)(z_{A}-z_i))] .
\end{eqnarray}

The above expression for the Green function and its ${\tilde s}$ dependence
contain the central result of this section. As the Green function 
enters directly into the production amplitude, we see that
the production amplitude in finite nuclei is necessarily a superposition of 
two separate resonance structures: the original, narrower
resonance peak due to decay outside the nucleus, and the
broader peak due decay inside the nucleus.
Note that the term $G_{in}$ does {\it not} have a pole singularity at  
${\tilde s} = M_R^{\ast 2}-iM_R^{\ast}\Gamma_R^{\ast}$ because the
residue of the pole, $C_{in}$, vanishes at this point. The Green function
thus has no pole corresponding to a medium modified resonance due
the finiteness of the nuclear medium. Nevertheless, for real values
of $\tilde{s}$, the amplitude does exhibit a Breit-Wigner structure.

The relative weights of the two components are given by the probability
amplitude for the decay inside and outside.
If it is known experimentally that one has detected decay products 
corresponding to decay of the resonance at some distance
from the target, only the narrow component will contribute. On the 
other hand, for an infinitely extended nucleus
$z_A\rightarrow\infty$ only the broad component is present with a
non-vanishing residue. 
For a finite nucleus, both contribute to the amplitude and interfere 
with each other in the cross section. As one is usually interested 
in the medium modified part,
both contributions must be carefully separated.
This situation is quite different from a coherent nuclear {\it formation}
reaction \cite{9} -  \cite{11}, where only the propagation of the
broadened resonance (or in medium component) is important.

For a finite nucleus with constant density we
can obtain an analytical expression for the forward production amplitude, 
$T$, by carrying out the 
integration in Eq. (\ref{eq:19}) also over the initial coordinate,
${\bf r}_i$. 
The result again shows separate contributions from decay
inside and outside of the nucleus:
\begin{equation}
\label{eq:25a}
T(p,0,{\tilde s})=
D_{in}(x,y) \cdot\left [{\tilde s} -M_R^{*2} +iM_R^* \Gamma_R^* \right ] ^{-1}
+D_{out}(x,y) \cdot\left [{\tilde s} -M_R^2 +iM_R \Gamma_R \right ] ^{-1} ~,
\end{equation}
where
\begin{equation}
\label{eq:25b}
D_{in,out}(x,y) = t(R\rightarrow X)t_0(s,0)\rho_0
\int\! d^2b\!\int_{-z_A(b)}^{z_A(b)} dz_i\,
C_{in,out}({\bf r}_i;{\bf p}-{\bf q};E_0) ~,
\end{equation}
and we have introduced the dimensionless variables
\begin{eqnarray}
\label{eq:43a}
&&x=q_z R_A = R_{A} {M^2-m_0^2\over 2|{\bf p}|}~, ~~~~
y=Q_R^{\ast}R_A
={M_R^{\ast 2}-iM_R^{\ast}\Gamma_R^{\ast}-m_a^2\over 2|{\bf p}|} R_{A}.
\end{eqnarray}
Defining a function
\begin{equation}
\label{eq:44b}
K(x)={3\over 2}\int_{0}^{1}\zeta\,d\zeta\,e^{-ix\zeta}=
{3\over 2}{(1+ix)e^{-ix}-1\over x^2} ~.
\end{equation}
we can express the coefficients $D_{in}$ and $D_{out}$ in the
following form
\begin{eqnarray}
\label{eq:46}
D_{in}(x,y) &=& 
{iA\,t_{0}(s,0)\,\ t(R\rightarrow X)\over 2p}
\left[{K(x)-K(-x)\over x}+{K(-x)-K(-x+2y)\over y}\right] ~,
\nonumber \\
D_{out}(x,y) &=& 
-{iA\,t_{0}(s,0)\,\ t(R\rightarrow X)\over 2p}
{K(-x)-K(-x+2y)\over y} ~.
\end{eqnarray}
In the discussion above, we did not take into account the contribution
of intermediate states with higher mass than the resonance R 
because of the damping due to the nuclear form factor
which takes place at intermediate energies.
Before showing examples for resonance production at intermediate 
energies, we first discuss the production at high energies. The expressions
will turn out to look very similar, even though the space--time picture
is entirely different. Asssuming that only one intermediate state, the
resonance R, contributes the expression for the amplitude will be identical
to Eq. (\ref{eq:46}). As we will consider this case in Ch. 4, the reader
may proceed directly to these results concerning the two component 
structure of the amplitude and skip the discussion of the space time picture
at high energies in the next chapter.
\section*{\bf\large 3. HIGH ENERGIES}
\indent

The discussion of resonance propagation in nuclei of the 
previous chapter cannot be applied at high energies. The
nonlocality of the amplitudes $t_R$ and $t_0$ must be taken into 
account. The interaction at high energies is dominated by
intermediate multiple particle production and the amplitudes  $t_R$ and $t_0$
are almost purely imaginary due to the dominance of multiparticle 
intermediate states. 
Under these circumstances the description of the high energy reaction 
in terms of reggeon exchanges with the target becomes most natural. 

It was already discussed by Gribov \cite{Gribov} that a crucial 
feature of high energy interactions are the multiparticle fluctuations
of the fast hadronic projectile, which are taken into account by reggeon
exchange (see Fig.3). Only the low momentum part of the fluctuation will
interact with the target. Two body processes, like elastic
scattering or diffractive dissociation, only arise as the shadow of
these multiparticle processes as required by the unitarity condition.
The lifetime of the fluctuations -- and thus also the
nonlocality of the interaction -- is proportional to the hadron energy. For the 
scattering of hadrons off nuclei it was shown {\it e.g.} in 
Refs.\cite{GribovA}, 
\cite{Koplik} and \cite{BKKS} that there is therefore a critical 
energy of order 
$E_{crit}\sim m_a(\mu R_A)$, above which the length of the fluctuation becomes 
larger than the nuclear size; here $m_a$ is the projectile mass and $\mu$ a 
hadronic scale parameter of the order of a few hundred MeV. 
Two effects become important above this energy. Coherent production of higher
mass states, $\sim (m_a + \mu)^2$, is not damped anymore by the 
nuclear form factor. Secondly, the nonlocality of the the interaction of a fast
hadron with a target nucleon becomes greater than the nuclear
dimension, $R_A$. This latter effect can be understood as follows.
In hadron -- nucleus scattering at high energies the hadron typically 
enters into a multiparticle state long before it enters the nucleus and 
reappears out of such a fluctuation  only far outside the 
nucleus. The space--time picture of resonance production as a
localized production and successive rescatterings of the resonance $R$ 
in nuclear matter, which we used in the previous Chapter, is therefore
not applicable anymore. It means that the sequential multiple scattering 
diagrams of Fig.2 -- called {\em planar} diagrams because of the topological 
structure of the upper part -- do not dominate anymore. Note that at high 
energies the wavy lines denoting the interactions with the nucleon stand 
for the exchange of one or more reggeons ({\it i.e.} multiparticle ladders). 
Intermediate multiparticle states correspond to cuts of these ladders.

While the intermediate coupling to multiparticle states through {\em sequential} 
processes in finite nuclei is negligibly small at high energies, a more 
complicated type of coupling to the multiparticle states becomes dominant. 
This is the {\it simultaneous} appearance of fluctuations. An example, a
{\em non-planar}  reggeon diagram, is shown in Fig.4.
The direct calculation of the contributions from such non-planar 
diagrams would be extremely difficult. Gribov \cite{GribovA} was able 
to include 
them in his discussion of high energy scattering and showed that surprisingly
the structure of the total scattering amplitude turned out to be analogous
to that obtained from the eikonal multiple scattering approach 
used in the previous chapter. 
We shall only outline briefly what Gribov's arguments \cite{GribovA}
for hadron--nucleus 
interactions at high energies imply for resonance production 
on nuclei. The derivation is based on the analytical properties of the
reggeon amplitudes as a function of the complex variables corresponding to the
invariant masses of intermediate states, $\tilde {s}_i=M_i^2$. This
analytic structure reflects the appearance of intermediate states and the 
nonlocality of the interaction.

For the coherent production of the resonance $R$ by a high--energy  
projectile $a$, we separate in the amplitude the part $A_{aR}$, which 
contains all reggeon exchange dynamics, indicated schematically
in Fig. 5:
\begin{eqnarray}
\label{eq:31}
T(p,q^2,\tilde {s}) &\sim & 
\sum_{n=1}^{\infty} \int\! d^2 b 
\!\int\!\cdots\!\int\!dz_n\,\ldots \,dz_1\,
\rho_A(z_n,b)\cdots\rho_A(z_0,b) \nonumber \\
&\times &
\int\! dq_{n-1,z}\ldots dq_{0z} 
\exp \{-i\sum_{i=0}^{n-1}q_{iz}(z_{i+1}-z_{i})\}
A_{aR}(p;q_{iz}) ~.
\end{eqnarray}
The integrations over the $z_i$ are longitudinally ordered, and 
the integrations over the $q_{iz}$ are carried out 
from $-\infty$ to $+\infty$ as required by the Feynman rules. 

The function $A_{aR}(p;q_{iz})$, the amplitude for a diffractive $a\rightarrow R$ transition through the interactions with $n$ target nucleons, can also be considered as a function of the variables $\tilde{s}_i$. This is possible since
the invariant masses $\tilde{s}_i$ of the $(n-1)$ intermediate diffractively produced states can in the eikonal limit be expressed 
in terms of longitudinal momentum transfers, $q_{iz}$, as was done 
in Chapter 2:
\begin{equation}
\label{eq:30}
\tilde {s}_i=(p-q_i)^2
\approx 2|{\bf p}| q_{iz}+m_a^2 ~,~~~i=0,\ldots,n-1 ~,
\end{equation}
where $m_a$ is the mass of the incident projectile $a$.

The important feature of the reggeon-exchange dynamics is a power-like decrease of amplitudes with increasing $\tilde{s}_i$ \cite{M}. It means that integrals over $\tilde{s}_i$ are convergent, and it is possible to modify integration contours in the complex $\tilde{s}_i$-plane by adding to them integrals over large semi-circles which are negligible.

We first discuss the analytical structure of the integrand of the simplest planar diagram, Fig.6a, and show that at high energies its contribution is negligibly small, as already argued above. 
On the positive real axis, we see in Fig.6a a pole singularity, corresponding to a possible stable single-particle state. 
The branch point corresponds to multiparticle states, contained 
in the internal structure of the exchanged reggeons (related to the
nonlocality of reggeon interactions).
The pole due to the resonance $R$ lies on the unphysical sheet. 
The presence of only right-hand-side but not left-hand-side singularities
is a specific property of planar diagrams.
There are also singularities due to the nuclear form factor in the $\tilde{s}_i$-plane. These singularities occur at $|q_z|\sim R_A^{-1}$
as can be seen by assuming, for example, a form factor of the structure 
$F_A\sim (1+q^2R_A^2)^{-1}$. At high energies, as $|{\bf p}|$ becomes 
asymptotically large, they thus correspond to far away singularities,
see Eq.(\ref{eq:30}), and can be neglected.
The nuclear formfactor singularities are not shown in the figure.

The solid line $C$ in Fig. 5a indicates the desired integral along the 
real axis, which we now want to rewrite. Combining the contribution from the contour $C$ and the contribution from the upper semi-circle (dashed), we obtain zero, since no singularities are enclosed 
by the contour. Due to the behavior of the reggeon exchange 
amplitude for large $\tilde{s}_i$, the contribution from the
semi-circle is negligible and thus the desired integral must be zero.
It means that at high energies the different singularities 
of the planar diagram cancel each other when one takes into account all intermediate multiparticle states. At energies lower than the critical one, 
the nuclear form factor is crucial and prevents this cancellation by damping large-mass contributions. (The form factor induced singularities are situated at small $\tilde{s}_i$ values in this case).

The situation is different for the non-planar graphs shown in Fig.4. There are now also singularities on the negative real axis, see Fig.6b. 
In this case, we can modify the integration contour combining the 
contributions from $C$ and the closed contour in the lower halfplane.
The parts along the negative 
axis cancel and we obtain a contribution from the contour enclosing 
the positive $\tilde {s}_i$-axis on the physical sheet.
By combining the integrand above and below the cut, only the discontinuity of the amplitude is left. Due to the unitarity relations this corresponds to the contribution of on-shell intermediate states.

In order to deal with physical intermediate states in constructing the 
total amplitude, the crucial point in the treatment by Gribov consists now 
in combining 
the absorptive parts of planar and non-planar contributions, {\it i.e.}
of diagrams with different topological structures. Thus the integral
reduces to the contribution from the absorptive parts in the $\tilde{s_i}$
variables and the final result is thus expressed in terms of
a sum or integration over {\em real} intermediate states, regardless of the
reaction mechanism (planar or non-planar). These states are 
labeled with momentum or mass values according to Eq.(\ref{eq:30}). 
The answer is therefore obtained in the form of multiple scattering 
between physical states, mediated by
reggeon exchange. In general, these 
amplitudes are nondiagonal, {\it i.e.} connect different states. 
The final formulas thus now include instead of amplitudes $t_0$ and
$t_R$ the matrix amplitude $\hat{t}$ which describes all possible
diffractive transitions between different states. Similarly, the effective
potential becomes a matrix, {\it e.g.} $\hat{U}(q)=\hat{t}\cdot F_A(q)$
in its momentum representation. It is convenient also to
introduce a diagonal matrix for the intermediate longitudinal
momenta for different states $\hat{Q}$ where
\begin{equation}
\label{eq:32a}
\hat{Q}_{cd}={M_c^2-m_a^2\over 2|{\bf p}|}\delta_{cd}~.
\end{equation}
It allows one to express the free Green function as a matrix analogue
of Eqs. (\ref{eq:5}) - (\ref{eq:6}):
\begin{eqnarray}
\label{eq:32}
\hat{G}_0({\bf r})&=&
e^{ipz}\hat{g}_0 (z)\,\theta (z)\,\delta^2 ({\bf b}) ~, \\
\hat{g}_0(z)&=&(2|{\bf p}|)^{-1}\exp(-i\hat{Q} z) ~.
\end{eqnarray}

The matrix amplitude describing scattering from the initial projectile to
a final hadronic state with nucleus staying in the ground state
has the form
\begin{eqnarray}
\label{eq:33}
\hat{T} ^{(n)} (p;b)&=&
i^{n-1}\int dz_n\,\ldots dz_1\,dz_0\,
\hat{U}(z_n,{\bf b})\hat{g}_0(z_n-z_{n-1})\times  
\nonumber \\
&&\times\hat{U}(z_{n-1},{\bf b})\cdots\hat{U}(z_1,{\bf b})
\hat{g}_0(z_1-z_0) \hat{U}_0(z_0,{\bf b}) ~,
\end{eqnarray}
where $\hat{U}(z,{\bf b})=\hat{t\cdot}\rho(z,{\bf b})$.
The index '0' on the  potential $U_0$ denotes initial interaction 
of the projectile. This incident channel may involve a photon,
which will needs not be taken into account as an intermediate state again.

It is convenient not to perform the last $z$ integration
but to define the operator functions 
$\hat{F}^{(n)}(z;b)$ and $\hat{F}(z;b)$, 
\begin{eqnarray}
\hat{T}^{(n)}(b)&=&\int_{-z_A}^{z_A}dz\,\hat{F}^{(n)}(z;b)~, \\
\hat{F}(z;b)&=&\sum_{n}\hat{F}^{(n)}(z;b) ~.
\end{eqnarray}
The function $\hat{F}(z;b)$ satisfies the Gribov integral equation
\cite{GribovA} 
\begin{equation}
\label{eq:34}
\hat{F}(z;b)=\hat{U}_0(z,{\bf b})
+i\int_{-z_A}^{z}dz_1\,\hat{U}(z,{\bf b})\,
\hat{g}_0(z-z_1)\,\hat{F}(z_1;b) ~.
\end{equation}

It is surprising that the final formulas obtained by Gribov have the same 
multiple scattering form as in the simple intermediate energy approach, extended to 
the whole set of intermediate states which can be produced coherently. However, 
the simple space--time interpretation of the interaction with the nucleus is 
lost. Thus, it would be wrong to interpret the final 
amplitude as the propagation of the resonance or another hadronic state through
the nucleus with {\em successive} rescatterings off the target nucleons. 
If the energy is higher than the critical one the correct picture corresponds
to {\em simultaneous} interactions mediated by reggeon exchange. 
The range of the interaction taking the initial projectile to its final state 
is typically larger than the nuclear dimension.

The general solution of Eq. (\ref{eq:34}) of course involves a complicated matrix problem. 
However, in the approximation of constant nuclear density $\rho_0$ the solution can be found using the method of 
Laplace transformations. Introducing Laplace transforms for functions $F(z)$ and $g_0 (z)$,
\begin{eqnarray}
\label{eq:lap}
\hat{{\cal F}}(\xi )&=&\int_{-z_A}^{\infty} e^{-\xi z}
\hat{F}(z) ~,\\
\hat{{\cal G}}_0 (\xi)&=&\int_{0}^{\infty} dz e^{-\xi z} 
\hat{g}_0 (z) =(\xi +i\hat{Q})^{-1}~,
\end{eqnarray}
we get in the case of semi-finite matter with constant density $\rho_0$ a simple algebraic matrix equation instead of integral equation (\ref{eq:34}) (we suppress
for the moment the dependence on $b$):
\begin{equation}
\hat{{\cal F}}(\xi)=\hat{U}_0/\xi+
i\hat{U}\hat{{\cal G}}_0\hat{{\cal F}}(\xi) ~,
\end{equation}
where $\hat{U}=\hat{t}\rho_0$.

It has the following solution
\begin{eqnarray}
\label{eq:36}
\hat{{\cal F}}(\xi)&=&{\exp{\xi z_A}\over \xi}
\left[ 1-i\hat{U} (\xi+i\hat{Q})^{-1}\right]^{-1}\,
\hat{U}_0 \nonumber\\
&=& {\exp{\xi z_A}\over \xi}
G(\xi)
\hat{U}_0 ~,
\end{eqnarray}
where
\begin{equation}
\hat{G}(\xi)=\left[\xi +i\hat{Q}^{\ast}\right]^{-1}~,
\end{equation}
with
\begin{equation}
\label{36a}
\hat{Q}^{\ast}=\hat{Q}-\hat{U} ~.
\end{equation}

The function $\hat{F}(z)$ can be expressed through the inverse
Laplace transformation
\begin{equation}
\label{eq:35a}
\hat{F}(z)={1\over 2\pi i}\int_{\uparrow}d\xi\,e^{\xi z}\,
\hat{{\cal F}}(\xi) ~,
\end{equation}
where the integration is performed along a contour which is
parallel to imaginary axis at a distance such that all singularities of the
integrand are on the left-hand side of it.

The expression for $\hat{{\cal F}}(\xi)$, Eq. (\ref{eq:36}), was derived in the
approximation of semi-infinite matter, $(-z_A,\infty)$. To get a result
valid for nuclear matter of finite size, $(-z_A,z_A)$, Eq.(\ref{eq:24}),
$$\rho_A(z;b)\propto \theta(z_A-|z|)~,$$ one has to multiply the Laplace
inversion of Eq. (\ref{eq:36}), $F(z)$, by $\theta(z_A-z)$ 
\begin{equation}
\langle  F(z) \rangle = F(z) \theta(z_A-z) ~,
\end{equation}
where we have denoted the quantities corresponding to the
finite-matter case with the brackets $\langle~\rangle$):

The general approach of Gribov is easily applied to the special case of resonance production on a nucleus. If one of the intermediate states is a resonance, 
one simply has to replace $M_c^2$ in the Green function, Eq. (\ref{eq:32}), by
$M_R^2-iM_R\Gamma_R$. 
To get the amplitude of resonance decay, the matrix
element $\{\hat{F}(z,b)\}_{Ra}$ should be convoluted with the free
Green function $\{\hat{G}_{0}({\bf r}_f-{\bf r})\}_{RR}$ and multiplied with
the vertex function $t(R\rightarrow X)$ introduced in the
previous chapter that describes the final decay of the resonance:
\begin{eqnarray}
\label{45}
T(p,q,\tilde{s})&=&\int d^2b\,e^{-i{\bf q}_{\perp}{\bf b}}
\hat{T}(\tilde{s};b) ~,\\
\label{46}
T(\tilde{s};b)&=&t_R(R\rightarrow X)
\int_{-z_A}^{z_A}dz \int_{z}^{\infty}dz_f
\left\{\hat{g_0}(z_f-z)\right\}_{RR}
\left\{\hat{F}(z,b)\right\}_{Ra}
 e^{iq_z z_f} ~,
\end{eqnarray}
where $q_z=(\tilde{s}-m_a^2)/2|{\bf p}|$. 
   
Note that the semi-infinite integral over $z_f$
can be seen as a complex Laplace transformation to a variable $\xi$,
related to momentum $q_z$, or invariant mass $\tilde{s}$,
\begin{equation}
\label{eq:xi}
\xi =  (\tilde{s} - {m_a}^2)/(2 i |{\bf p}|)
\end{equation}
(compare Eq. (\ref{46}) and (\ref{eq:lap})).
Therefore the Laplace-transform 
method is very 
convenient for the calculation of the resonance mass distribution.
The expression (\ref{46}) has a form of a convolution of two functions, 
which corresponds to a product of two Laplace transforms in variable 
$\xi$ :
\begin{equation}
\label{54}
T(\tilde{s};b)=t_R(R\rightarrow X)
\langle {\cal F}(\xi ;b)\rangle_{aR}
\left\{{{\cal G}}_0(\xi)\right\}_{RR}  ~.
\end{equation}

To get the Laplace transform for the truncated function 
$\langle F(z)\rangle$, one should convolute Eq. (\ref{eq:36}) with the Laplace 
transform $\theta_A(\xi)$ of the step function $\theta(z_A-|z|)$:
\begin{eqnarray}
\label{eq:38}
\langle\hat{{\cal F}}(\xi)\rangle&=&
\int_{\uparrow}d\xi^{\prime}\,
\hat{{\cal F}}(\xi^{\prime})\,\theta_A(\xi-\xi^{\prime}) ~,\\
\theta_A(\xi)&=&
{\exp(z_A\xi)-\exp(-z_A\xi)\over \xi} ~.
\end{eqnarray}
For (semi-)infinite nuclear matter the amplitude can be represented as a
sum of poles, 
\begin{equation}
\label{eq:39}
\hat{{\cal F}}(\xi)=\sum_{i}{a_i\over \xi -\xi_i} ~,
\end{equation}
and the finite-matter amplitude then has the form
\begin{equation}
\label{eq:40}
\langle\hat{{\cal F}}(\xi)\rangle=
\sum_{i} a_i\,\theta_A(\xi-\xi_i) ~.
\end{equation}

The general formalism in the multichannel case can be written in a
compact fashion using functions of a
matrix argument. An essential point is the ordering of
different matrix factors.  From the pole structure of the function 
$\hat{{\cal F}}(\xi)$
one gets by means of Eqs.(\ref{54}), (\ref{eq:38}), and (\ref{eq:36})  
\begin{eqnarray}
\label{101}
T(\tilde{s};b)&=&t_R(R\rightarrow X)
(\xi +i\hat{Q})^{-1}\left\langle 
{\exp (z_A\xi)\over \xi }(\xi+i\hat{Q})
(\xi+i\hat{Q}^{\ast})^{-1}\right\rangle \hat{U}_0 
\nonumber \\
&=&
\left[
\hat{Q}(\hat{Q}^{\ast})^{-1}\,{\exp(z_A\xi)-\exp(-z_A\xi)\over \xi}+\right.
\nonumber \\ 
&&\left. +(\hat{Q}^{\ast}-\hat{Q}) (\xi+i\hat{Q}^{\ast})^{-1}
(\hat{Q}^{\ast})^{-1}\,
\{\exp(z_A\xi)-\exp[-z_A(\xi+2i\hat{Q}^{\ast})]\}
\right]
~.
\end{eqnarray}

The important feature of hadron--nucleus interactions at
high--energies, the possibility for the initial hadron to
convert in the course of rescatterings into different hadronic
states,is determined by the matrix $\hat{U}$ contained in the
propagator $\hat{G}$.
Therefore it is necessary to know the structure of the effective potential, 
$\hat{U} = \hat{t} \rho$, and thus of the elementary 
diffractive matrix amplitude $\hat{t}$ through diffractive dissociation
processes. At present, this information is essentially limited
to reactions involving at least one stable hadronic state.
It is known that for stable particles, such as protons or pions,
the nondiagonal elements are much smaller than the diagonal
ones \cite{kaid}. There are some arguments for the magnitude of diagonal
amplitudes for unstable hadrons: for $\rho$ mesons they are of the same order 
as for pions,
and for $\Delta$ isobars comparable to that for nucleons. 
Resonance production on nuclei can provide
further information on the structure of the hadron--nucleon diffraction
matrix.

Some conclusions about the structure of the nuclear diffraction amplitude,
 $T$, can be obtained  at high energies when the free part 
of the Green function, $(\tilde{s} - M_R^2 + i M_R \Gamma_R) / 2 |{\bf p}|$,
 may be neglected compared to its density dependent part, $\hat{U}$:
\begin{equation}
\hat{G}(\xi)=\left[\xi +i\hat{Q}^{\ast}\right]^{-1}
\approx \hat{U}^{-1} ~.
\end{equation}
As a result, the amplitude is proportional to
\begin{equation}
\hat{T}(\tilde{s},b)\propto (\xi+i\hat{Q})\hat{U}^{-1}\hat{U}_0 ~.
\end{equation}
If the projectile particle is a hadron, {\it i.e.} $\hat{U}_0=\hat{U}$, the
resulting $\hat{T}$ matrix is diagonal. This takes place
even if there are non-diagonal hadron--nucleon matrix elements
which are not small. 
Note that while the Green function in nuclear matter is not diagonal
(as is $\hat{U}^{-1}$), the $\hat{T}$ matrix becomes 
diagonal due to a cancellation between the resonance production
amplitude and the terms induced by subsequent rescatterings.

To be more precise, not the matrix elements,
but the eigenvalues of the matrix $U$ should be large compared to
other terms in the Green function denominator to lead to a diagonal
total amplitude. If one of eigenvalues is small for some reason, the
situation can be quite different. 
The interaction with the
nucleus looks much simpler in terms of eigenstates of the matrix
amplitude of hadronic scattering. The initial hadron can be
represented as a linear combination of diagonal states
propagating through the nucleus. These diagonal states have, in
principle, different probabilities to be absorbed in nuclear
matter. As a result, the nucleus works as a
filter that only lets hadronic states with the smallest cross section
pass.

In the case of photoproduction, the incident photon can be
represented as a superposition of different hadronic states. In
the vector dominance model the main contribution comes from
low-lying vector mesons ($\rho,\omega,\phi$). If one of these
resonances is detected in the final state, the non-diagonal
transitions in intermediate states are not important and the
amplitude can be represented in the form of Eq.(\ref{eq:25a}). For
photoproduction of higher resonances, such as the
$\rho^{\prime}$, several mechanisms can contribute:
$<\gamma|\rho><\rho|T|\rho^{\prime}>$ and
$<\gamma|\rho^{\prime}><\rho\prime|T|\rho^{\prime}>$. They are
related to the admixtures of $\rho$ and $\rho^{\prime}$ in the
photon wave function, respectively.  Although the admixture of
$\rho^{\prime}$ is small, the second mechanism is diagonal in
nuclear transition and therefore increases faster with $A$ as
compared to the first one.  Thus, the coherent photoproduction of
higher vector mesons on a nuclear target can provide useful
information on the hadronic structure of the photon wave
function.

\subsection*{Single channel case}

We now consider the general amplitude for the case when only a single 
state, the resonance $R$, of all intermediate states is essential. 
In this case Eq.(\ref{101}) simplifies since it involves only scalar quantities
$Q_R$ and $Q_R^{\ast}$ instead of matrices $\hat{Q}$ and $\hat{Q^{\ast}}$. 
\begin{equation}
\label{eq:41}
T(\tilde{s};b)=t_R(R\rightarrow X)\,{t_0(s,0)\,\rho_0 \over 
[1-it_R\,\rho_0\, (\xi+iQ_R)^{-1}]} {\exp(z_A\xi)\over \xi}~.
\end{equation}

After integration over the impact parameter $b$ we get for forward 
resonance production, {\it i.e.} when $q^2 = -q_{z_,min}^2$ ,
\begin{eqnarray}
\label{eq:45}
&&T(p,-q_{z_,min}^2,\tilde{s})
=
\int\! d^2b\, T(\tilde{s};b)
\nonumber \\ 
&&={4\over 3}\pi R_A^3\,\rho_0\,t_0(s,0)\, {i\over x}\left[ 
{K(x)-K(-x+2y)\over \xi+iQ_R^{\ast}} 
+{Q_R\over Q_R^{\ast}}{K(-x)-K(-x+2y)\over \xi+iQ_R}
\right] .
\end{eqnarray}
The function $K(x)$ and the kinematical variables $x$ and $y$ are given
in Eqs. (\ref{eq:44b}) and (\ref{eq:43a}), respectively.
This expression can be 
transformed to yield the same amplitude as in Ch. 2, Eq. (\ref{eq:25a}),
which was obtained for lower energies. It illustrates that the high energy formulas are a smooth continuation
of the intermediate energy expressions, even though the underlying
space-time picture is completely different: sequential 'planar' 
contributions dominate at lower energies and 'non-planar' mechanisms at
high energies.
\section*{\bf\large 4. EXAMPLES}
\indent

As was shown in the preceeding Chapter, we can simply use the single channel expression with its two component structure over a wider range of energies
as long as it is reasonable to neglect the coupling to other channels.  
We now show some implications of the presence of two resonance components
for the invariant mass distribution of the decay products. This is the
natural aspect to study if one is interested in the behavior of the
resonance in the nuclear medium.
We consider a situation with a high momentum $|{\bf p}|$ where  
the production is not damped by the nuclear formfactor, since 
$q_{min} \approx (\tilde{s} - {m_a}^2)/2|\bf{p}|^{-1}$. This implies 
that the broadening of the resonance is large compared to its free width.  
Then in the mass distribution the narrower peak from the decay in the
vacuum must compete with the broad in-medium contribution.
The interference between these two terms with a Breit--Wigner structure
will be important for the mass distribution. 
In contrast to the usual non-resonant background, the form and 
relative phase of the in medium 'background' here are determined by 
the resonance interaction with the medium.  
(Certainly there will also be a background of non-resonant origin in
the actual measurement; we have neglected it here). 

We present as an illustration the application of
the two--component formula, Eq. (\ref{eq:25a}), to coherent photoproduction
of a $\rho$ meson. In Fig.7 we show the mass spectra of the $e^+e^-$
pair from $\rho$ decay produced at $\theta =0^0$ in the reaction
\begin{equation}
\label{eq:28a}
\gamma A\rightarrow\rho A\rightarrow (e^+ e^-)A 
\end{equation}
at laboratory momenta $p_{lab}$ of $2$ and $5$ GeV and for finite nuclei 
with $A=50$ and $A=200$.
The cross section $d\sigma /d{\tilde s}$ was calculated using the amplitude
from Eq. (\ref{eq:25a}) for a constant nuclear density, Eq.(\ref{eq:24}),
with $\rho_0=(4/3\pi R_A^3)^{-1}~,~~R_A=1.12 A^{1/3}~fm$. 
The forward $\rho N$ scattering amplitude was assumed to be purely imaginary and
$\sigma_{\rho N}^{tot}$ was taken to be 20 mb.  
The solid and dotted curves show the contributions of the free and
in medium components, respectively. The bold curves represent the total
contribution, which includes the interference of the components.
The narrow component has a Breit--Wigner structure, which is
distorted due to the dependence of the nuclear form
factor, $F_{A}(q_{min})$, on ${\tilde s}$ through $q_{min}$. To partially 
remove this effect, we have scaled the cross section by a factor 
$\tilde{s}^2$. The distortion due to the presence of two components in the
amplitude is especially visible for the lower energy, $p = 2 GeV$ and
the heavier
nucleus: The in-medium contribution separately can be seen as a 
broad resonance peak, where the form factor distortion is much
more evident. The
interference between broad and narrow components is dramatic at 2 GeV
and completely changes the form of the mass distribution. It
is largely constructive at $\sqrt{{\tilde s}}<M_R$ and 
destructive at $\sqrt{{\tilde s}}>M_R$.
Due to the strong interference the form of
the free $\rho$-peak becomes completely distorted and asymmetric at $A=50$
and even develops two minima near 0.7 and 0.9 GeV for $A = 200$. 
The more complicated picture for heavy nuclei is again related
to the presence of the nuclear formfactor, $F_{A}(q_{min})$, which is 
more rapidly varying for heavier nuclei. We have tested that the
general interference features are not an artifact of the sharp edge of the 
nuclear density distribution assumed in Eqs. (\ref{eq:8}) and (\ref{eq:14})
by repeating the calculations with a smoothly varying density.

The width of the broad component increases linearly with $p$. 
At 5 GeV it appears only as a small and broad background. The 
interference of the broad and narrow component changes the form of 
the $\rho$-peak  again, but clearly less than at the lower energy.

We see that at both energies the broad component becomes more 
important as the nuclear mass number, $A$, increases. The
different mass dependence can be made explicit in the
following way. We consider a situation when the momentum, $|{\bf p}|$, is
high and the broadening of the resonance is much larger than
its free width, $\Gamma_R^{\ast} \approx \rho_0 \sigma_R / 2 M_R \gg 
\Gamma_R$. The in-medium contribution then only represents a 
broad background to the pronounced narrow free peak.
In the vicinity of free peak we have
\begin{eqnarray}
\nonumber
x\approx M^2\,R_A/ |{\bf p}|\ll 1,\\ 
y\approx -~i~R_A~\rho_0~\sigma_R/2,~|y|\gg 1~.
\end{eqnarray}
Expanding $K(x)\approx 3/4-ix/2$ and $K(x+2y)\ll K(x)$, one
obtains for the two terms in Eq. (\ref{eq:45})
\begin{equation}
\label{eq:47}
\langle {\cal T}_{aR}(\xi)\rangle\approx
f_0\left[ \left(A-{2\pi R_A^2\over\sigma_R} 
\right){1\over \xi+iQ_R^{\ast}}+
{2\pi R_A^2\over\sigma_R}{1\over \xi+iQ_R}\right] ~.
\end{equation}
Since ${R_A}^2 \sim A^{2/3}$, this shows that the medium modified
part grows faster with $A$ than the free resonance contribution.

In the above consideration it is easy to estimate the effect
of correlations between nucleons \cite{GribovA}. The most
important are two--body correlations which can be be incorporated by 
simply multiplying the Green function $g_0(z_i-z_{i-1})$ by a correlation
function. This function $\chi(z_i-z_{i-1})$ is defined to be unity if
correlations are absent and we assume for an estimate a simple behavior
corresponding to a hard core  with radius $r_{corr}$,
\begin{equation}
\label{eq:48}
\chi(z)=\theta (|z|-r_{corr}) ~. 
\end{equation}
Its Laplace transform is
\begin{equation}\label{eq:49}
\chi(\xi)={\exp(-\xi a)\over \xi}~.
\end{equation}
This yields a resonance denominator in (\ref{eq:41}) of the form
\begin{equation}
\label{eq:50}
\xi+iQ_R-it_R\rho_0\exp{-i(\xi+iQ_R)r_{corr}}\approx
(1-\epsilon)
(\xi+iQ_R+{1\over 2}\rho_0\sigma_R (1+\epsilon))~,
\end{equation}
where $\epsilon={1\over 2}\rho_0\sigma_R r_{corr}$.
Thus, the inclusion of correlations results in a modification of the 
resonance width of second-order in the density, 
$Q_R^{\ast}\rightarrow Q_R^{\ast}+{1\over
2}\rho_0\sigma_R\epsilon$. 
For typical values of the parameters entering into $\epsilon$, the correction
to the induced width is not very large: For $\sigma_R\approx 20~mb$ and
$r_{corr}\approx 0.2 fm$ one obtains $~\epsilon\approx 0.2$.

In summary, the amplitude for the production of hadronic resonances 
on nuclear targets contains two types of components. They correspond 
to the propagation and decay of the resonance outside and 
inside nucleus. The narrow component is characterized by the free
values of resonance mass and width. The broad 'in-medium' component 
does not have a true pole in the invariant mass; it has an 
approximate Breit--Wigner form with parameters
depending on  the nuclear density and the resonance--nucleon cross
sections.
Both components have a different $A$ and energy dependence. In
the limit of large $A$, the ratio $D_{in}/D_{out}$ of the broad
and narrow components is proportional to the nuclear size
$A^{1/3}$.  As the width of the broad component increases
linearly with energy, the inside/outside ratio at a fixed
invariant mass decreases as $1/p$. Thus, the most interesting
region for the investigation of 'in-medium' effects may be the
region of intermediate energies, where there is only moderate
damping by the nuclear form factor and the inside to outside
ratio is not suppressed. 
As this ratio is proportional to the fraction of time the
resonance spends inside the nucleus, it will be difficult to
establish nuclear effects for the narrow, long lived resonances
like $J/\psi$.
This makes it hard to extract the 'in-medium' parameters from 
such experiments in a simple fashion.

\section*{\bf 5
. Summary and Conclusions}
 
Experiments on nuclei are an important way to study the space-time
picture of elementary interactions. Through the interaction with the
nuclear medium one has a way of measuring the development of the
reaction in units of typical nuclear length and 
time scales. Many recent experiments concern the behavior
of unstable elementary particles - resonances - in nuclei. 
Ideally one would like to extract as directly as possible 
the 'in medium' properties of the resonance, {\it e.g.} the medium 
modified resonance parameters of mass and width. They can then
be compared to models - be it on the level of quarks or hadrons -
for the internal structure of the resonance and its interaction
with the target nucleons. In this paper we have looked specifically 
at the {\it production} of resonances. Examples are the production
of $\rho$ or $ J/ \psi$ on nuclei. For simplicity, we
left out final state interactions of the decay products 
in our more qualitative discussion; this is appropriate
for the decay of the above resonances into {\it e.g.} an $e^+ e^-$ 
pair. 

Clearly the relevant kinematical variable for such a study
is the invariant mass of the resonance, $\tilde{s}$, which enters into 
the resonance amplitude and governs the resonant shape of the
cross section. However, in general the connection between the 
incident energy, the variable at our disposal in an experiment, 
and the invariant mass $\tilde{s}$ 
of the resonance in the nucleus  is not direct in a high energy
collision: the nucleus will break 
up and different fragments will carry of part of the energy, with only
a variable fraction left for the resonant state we detect in the end. 
However, in a {\it coherent} process, where the nucleus returns to its 
ground state, the nuclear formfactor restricts the momentum transfer 
to the target and energy and invariant mass are closely related. That is 
why we discussed coherent production processes only.

In a production process, the resonance can decay inside and outside
of the nucleus. The relative contribution of these two possibilities
depends {\it e.g.} on the momentum of the resonance, the size of 
the nucleus and the resonance lifetime. Both possibilites show up
in the total amplitude: the decay outside through a component 
with the free resonance parameters and an additional medium-modified 
component for the decay inside the nucleus. 
The medium modifications can be described - in the simplest picture - 
as a shift 
and broadening of the resonance due to interactions with the nuclear
medium. As we saw in some illustrative examples, the background
contribution drops off fast with energy and we mainly see the free
resonance at high energies. This is intuitively clear since the faster
the resonance, the smaller its chance for an interaction with the
nucleus. A similar qualitative statement also holds for the dependence
on nuclear size: the importance of the medium-modified resonance
is greater the larger the nucleus. Thus in order to get at medium effects, 
it is best to stay at intermediate energies and chose a heavy
nuclear target. 

It is legitimate to talk about the medium-modified component of
the resonance in the production amplitude. However, it
must be stressed that for a finite nucleus the amplitude doesn't 
develop a pole corresponding to an 'in-medium' state. While there is 
a resonance denominator of the type $\tilde{s}-M_{R}^{\ast
2}+iM_{R}^{\ast}\Gamma_{R}^{\ast}$, where $M_{R}^{\ast}$ and
$\Gamma_{R}^{\ast}$ are the medium modified resonance parameters,
the amplitude has no pole because, as we showed, the residue
of the amplitude at
$\tilde{s}=M_{R}^{*2}+iM_{R}^{\ast}\Gamma_{R}^{\ast}$ is zero. 
There is only a pole in $\tilde{s}$ due to the free resonance,
{\it i.e.} a $(\tilde{s}-M_{R}^{2}+iM_{R}\Gamma_{R})^{-1}$
contribution with a non-vanishing residue. The reason behind this
is clear: we detect the resonance through its decay products far
from the target. 
Singularities in the $S$-matrix can thus not be due to 
the interaction in a limited space--time region, as in general 
the S - matrix has only poles corresponding to asymptotic states.
A simple illustration of this statement was shown by considering the
production of the resonance in an infinite nuclear medium,
$R_{A}\rightarrow \infty$. In that case the in-medium resonance
never gets out and a true pole does develop.
The above general conclusions were first derived in the standard
multiple scattering formalism, the eikonal description, and our
results can intuitively be understood in this picture. The
statements were then shown to apply at intermediate as well as high energies.
However, while the final formulas don't change, we must at high energies
radically change our space--time picture of the reaction, 
analogous to the findings by Gribov for the
scattering of a stable projectile from a nucleus. 
Multi-component intermediate hadronic
states, instead of a single hadron, propagate. These components
interact {\it simultaneously} with the target, instead of sequentially 
as is the case at lower energies. In the diagrammatic language, the 
sequential multiple scattering can be represented by
{\it planar} diagrams. The new element entering into the
description at high energies, above the critical energy, are {\it non-planar}
diagrams with an entirely different singularity structure. Rather
than dealing with this new contribution separately, it was discussed
how the consideration of planar and non-planar diagrams together,
grouped in such a way that the total contributions of real
intermediate states with a definite mass enter, leads to
a matrix expression with a 'multiple scattering' structure that is very 
similar to the amplitude at
$E<E_{crit}$. This result is quite surprising given the entirely
different underlying space-time picture.

We have seen that in coherent production on a finite nucleus
the medium-modified resonance amplitude is necessarily accompanied
by a free resonance contribution, which coherently
interferes with it in the expression for the cross section. In a
production process, we thus cannot directly measure an in-medium
resonance shape. The medium-modified component plays a significant role 
mainly at relatively low energies and for heavy targets, where due
to its interference with the free amplitude it can lead
to a significant change in the overall ${\tilde s}$ dependence of the 
cross section.
A qualitative estimate showed that
nuclear correlations will only have a moderate influence
on the distribution of the resonance decay products.
At high energies and on lighter targets, the in-medium
contribution quickly becomes a small, broad background that is
difficult to identify uniquely. What we see at high energies is
mainly the peak due to the free resonance, determined by the
vacuum parameters. At these energies, we also must be careful in our 
interpretation of the process, as a
totally different space--time picture applies than at lower
energies.

{\bf Acknowledgement.} The authors are indebted to A.B. Kaidalov for useful discussions and friendly criticism. L.K. thanks T.E.O. Ericson for stimulating
discussions with. K.B. and L.K. acknowledge support from grant J74100 of the
International Science Foundation, the Russian government and the Russian Fund 
of Basic Research. M.I.K. was also supported by a grant from the Alexander 
von Humboldt-Stiftung and by RFBR Grant 94-03068. The work of J.K. is part 
of the research program of the Foundation for Fundamental Research of Matter (FOM) and the National Organisation for Scientific Research (NWO). The collaboration was made possible by a grant from NWO as well as by INTAS 
grants 93-0023 and 93-0079.

\small{

}
\end{document}